\documentclass[journal]{IEEEtran}

\usepackage{amsfonts}
\usepackage{amsmath}
\usepackage{amssymb}
\usepackage{mathabx}
\usepackage{mathrsfs}
\usepackage{tensor}
\usepackage{esint}
\usepackage{tikz}
\usepackage{multirow}
\usepackage{makecell}
\usepackage{textcomp}
\usepackage{amsthm}
\usepackage{subfigure}
\usepackage{graphicx}
\usepackage{stackengine}
\usepackage{balance}
\usepackage{cite}
\usepackage{url}
\usepackage{adjustbox} 

\usepackage[caption=false,font=footnotesize,labelfont=sf,textfont=sf]{subfig}
\theoremstyle{remark}

\begin{document}

\title{ Low-Rank Augmented Implicit Neural Representation for Unsupervised High-Dimensional Quantitative MRI Reconstruction

}

\author{Haonan Zhang, Guoyan Lao, Yuyao Zhang, Hongjiang Wei,~\IEEEmembership{Member,~IEEE,}

\thanks{This work was supported by the National Key Research and Development Program of China under Grant 2024YFC2421100, the National Natural Science Foundation of China under Grant 62471296, Grant 62071299, the SJTU Trans-med Awards Research under Grant STAR 20220103 and Grant YG2023LC02. (Corresponding author: Hongjiang Wei.)} 

\thanks{Haonan Zhang and Guoyan Lao are with the School of Biomedical Engineering, Shanghai Jiao Tong University, Shanghai 200240, China (e-mail: haonanzhang@sjtu.edu.cn; guoyan.lao@sjtu.edu.cn).}

\thanks{Yuyao Zhang is with the School of Information Science and Technology, ShanghaiTech University, Shanghai 201210, China (e-mail: zhangyy8@shanghaitech.edu.cn).}

\thanks{Hongjiang Wei is with the School of Biomedical Engineering and the National Engineering Research Center of Advanced Magnetic Resonance Technologies for Diagnosis and Therapy (NERCAMRT), Shanghai Jiao Tong University, Shanghai 200240, China (e-mail: hongjiang.wei@sjtu.edu.cn).}}

\markboth{Journal of \LaTeX\ Class Files,~Vol.~14, No.~8, August~2021}%
{Shell \MakeLowercase{\textit{et al.}}: A Sample Article Using IEEEtran.cls for IEEE Journals}

\IEEEpubid{0000--0000/00\$00.00~\copyright~2025 IEEE}

\maketitle

\begin{abstract}
Quantitative magnetic resonance imaging (qMRI) provides tissue-specific parameters vital for clinical diagnosis. Although simultaneous multi-parametric qMRI (MP-qMRI) technologies enhance imaging efficiency, robustly reconstructing qMRI from highly undersampled, high-dimensional measurements remains a significant challenge. This difficulty arises primarily because current reconstruction methods that rely solely on a single prior or physics-informed model to solve the highly ill-posed inverse problem, which often leads to suboptimal results. To overcome this limitation, we propose LoREIN, a novel unsupervised and dual-prior-integrated framework for accelerated 3D MP-qMRI reconstruction. Technically, LoREIN incorporates both low-rank prior and continuity prior via low-rank representation (LRR) and implicit neural representation (INR), respectively, to enhance reconstruction fidelity. The powerful continuous representation of INR enables the estimation of optimal spatial bases within the low-rank subspace, facilitating high-fidelity reconstruction of weighted images. Simultaneously, the predicted multi-contrast weighted images provide essential structural and quantitative guidance, further enhancing the reconstruction accuracy of quantitative parameter maps. Furthermore, our work introduces a zero-shot learning paradigm with broad potential in complex spatiotemporal and high-dimensional image reconstruction tasks, further advancing the field of medical imaging.
\end{abstract}

\section{Introduction}
\label{sec:Introduction}
\IEEEPARstart{Q}{uantitative} magnetic resonance imaging (qMRI) provides tissue parameters such as spin-lattice relaxation time ($T_1$), spin-spin relaxation time ($T_2$), water molecule diffusivity, and tissue magnetic susceptibility \cite{1}, offering unique physical insights into brain tissue and enabling detailed characterization of its microstructure. Compared to single-parametric qMRI, multi-parametric qMRI (MP-qMRI) provides a more comprehensive view and has become essential in the diagnosis of healthy and diseased brain tissue by quantifying iron, myelination and cell membranes \cite{disease_1, disease_2, disease_3}. However, its clinical applicability is limited, mainly due to the need for multiple scans with different sequences, increasing the scan time and the risk of image misalignment. To address this, simultaneous MP-qMRI techniques have been developed \cite{metere2017simultaneous,5_MRF,christodoulou2018magnetic,6_EPTI,zhang2022blip,sun2020extracting, wang2022free}, such as magnetic resonance fingerprinting (MRF) \cite{5_MRF,liao2024high}, echo planar time-resolved imaging (EPTI) \cite{6_EPTI}, and MR multitasking \cite{chen2018strategically,7_MR_tasking}. Although these techniques enable the simultaneous measurement of different contrast information, the large number of captured temporal dimensions, ranging from dozens to over a thousand, necessitates accelerated acquisition to ensure practically feasible scan time. Modern accelerated MRI techniques primarily employ undersampled k-space acquisitions combined with advanced algorithms to estimate missing samples based on the sampled points. Therefore, a robust reconstruction method that can accurately produce multi-contrast weighted images and multi-parametric maps from sparse high-dimensional measurements is crucial for accelerated MP-qMRI.

Reconstructing multi-parametric maps from undersampled k-space data presents a challenging ill-posed inverse problem. Conventional reconstruction approaches primarily fall into two categories: dictionary-matching \cite{ma2013magnetic,8_dic_match} and low-rank tensor (LRT)-based methods \cite{9_LRT,he2016accelerated,cao2022optimized}. Dictionary-matching approaches estimate tissue-property maps by matching observed signals with precomputed templates in a dictionary. However, they face significant limitations as computational complexity grows exponentially with increasing temporal dimensions and resolution requirements, rendering them impractical for high-dimensional imaging scenarios. In contrast, LRT-based methods employ low-rank tensor decomposition to efficiently separate spatial features from spatiotemporal k-space data and reconstruct weighted images for parametric maps fitting through sequence-specific signal modeling. While the dimensionality reduction inherent in LRT  approaches simplifies the inverse problem, the estimation of  spatial bases becomes increasingly challenging and unstable at higher data dimensionalities and acceleration rates.

\IEEEpubidadjcol
Deep learning has emerged as a powerful paradigm for accelerated MRI reconstruction, demonstrating remarkable success in various applications \cite{wang2016accelerating,wang2022dimension,liang2020deep,cao2024high}. For MP-qMRI reconstruction, supervised deep learning approaches  
\cite{10_machine_learing,gao2021accelerating,cheng2022high,li2023supermap,zimmermann2024pinqi,lu2025acceleration} have shown particular promise by learning end-to-end non-linear mappings between undersampled multi-contrast MR images and tissue parameter maps. However, they typically require large, high-quality training datasets that are difficult to collect in the high-dimensional imaging scenarios, and they typically exhibit limited generalization across different acceleration factors or acquisition sequences. Unsupervised deep learning strategies offer an attractive alternative by directly reconstructing images from undersampled measurements without requiring external training data. Joint-MAPLE \cite{12_joint_maple}, for instance, employs an unsupervised convolutional neural network (CNN) to jointly estimate weighted images and optimize quantitative parameter maps from k-space data. Although this method integrates CNN with the unrolled physical model, 
its performance degrades with complex temporal variations due to insufficient utilization of measurement redundancy. Similarly, Zero-DeepSub \cite{zero_deepsub} combines CNN with the unfolded low-rank subspace model and simplifies the inverse problem to effectively restore weighted images. However, both approaches require further partitioning of already acquired undersampled k-space data, which compromises reconstruction stability and structural fidelity, particularly at high acceleration factors.

Recently, implicit neural representation (INR) has gained traction as a promising unsupervised learning paradigm for solving the inverse problem in medical image reconstruction. INR models medical images as continuous functions through a multi-layer perceptron (MLP) that map spatial coordinates to image intensities. By employing advanced coordinates encoding techniques \cite{15_encode_INR,21_5_fourier_encode,21_radial_encode}, INR can effectively capture high-frequency image details while overcoming the low-frequency bias inherent in CNNs. Moreover, INR's continuous representation provides a natural framework for incorporating physics-informed models in medical imaging \cite{13_INR_f_mi,14_INR_f_mi,16_encode_INR, chen2023cunerf, huang2024subspace}. However, current INR-based methods are limited by their reliance on single prior or model, restricting their ability to handle complex high-dimensional data. The recently proposed SUMMIT framework \cite{22_lao_singal_model} for MP-qMRI reconstruction, for example, utilizes only the Bloch signal equation to relate parametric maps to acquired k-space data, failing to exploit spatial structural correlation across temporal measurements. Consequently, SUMMIT struggles with artifact suppression and structural preservation at high acceleration factors, primarily due to its inadequate modeling of spatial redundancy in high-dimensional measurements.

\begin{figure}[!t]
    \centering
    \includegraphics[width=\columnwidth]{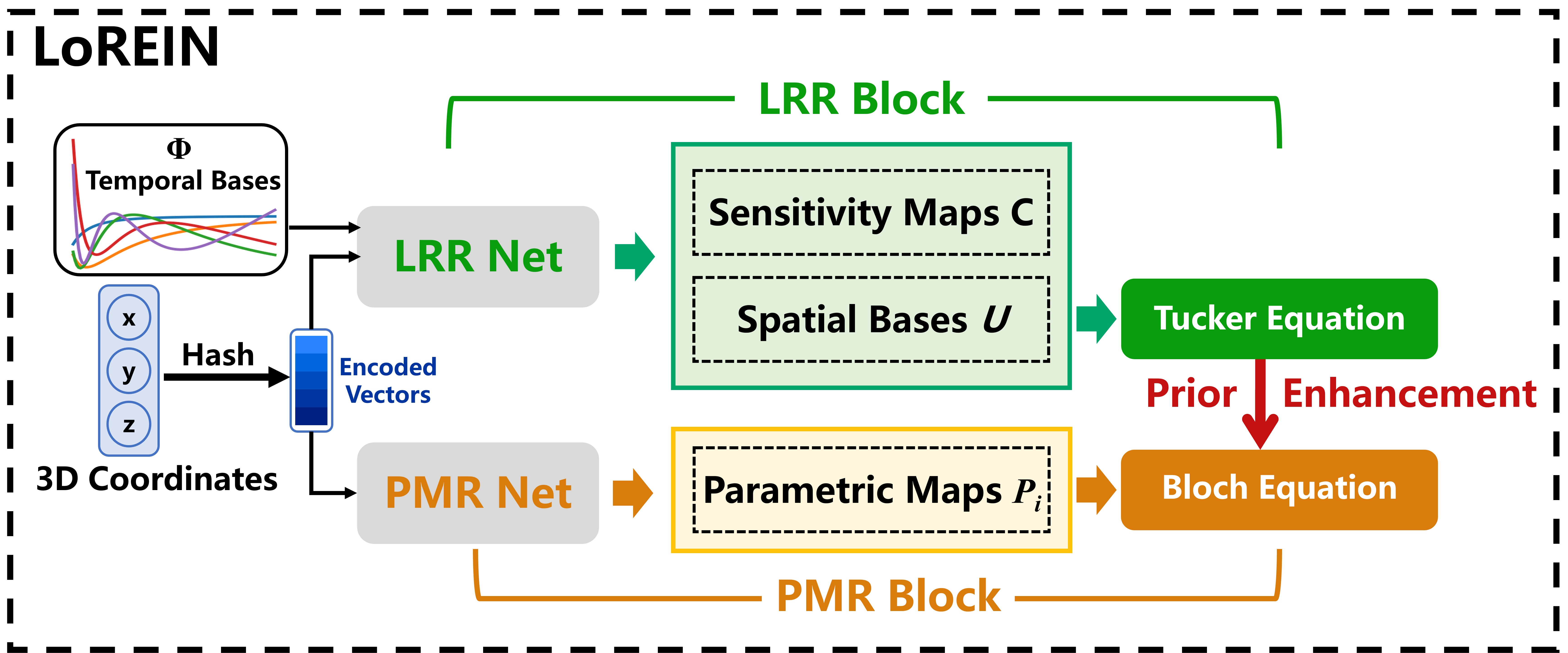}
    \caption{Simplified pipeline of proposed framework. LoREIN integrates two parallel blocks: the LRR block and the PMR block, both taking encoded vectors from 3D coordinates as input. The globally optimal spatial features of multi-contrast k-space are reconstructed through the dual-prior-driven LRR net, and the LRR-predicted feature guidance is transferred into the PMR block based on two signal models. Finally, dual models and priors are embedded in the unified optimization framework to achieve accelerated MP-qMRI reconstruction.}
    \label{fig_pipeline}
\end{figure}

In this study, we introduce LoREIN, a novel unsupervised framework for accelerated 3D MP-qMRI reconstruction (Fig. \ref{fig_pipeline}). Unlike existing approaches, LoREIN uniquely integrates two priors: low-rank and continuity to achieve robust reconstruction. The framework comprises two core modules: the Low-Rank Representation (LRR) block, which models spatial bases and coil sensitivity maps using coordinate-based MLPs, coupled with a specialized CNN for spatial bases refinement; and the Parametric Map Reconstruction (PMR) block, which represents multi-parametric maps as continuous functions of spatial locations. LoREIN jointly optimizes both modules using  two physics-based signal models, enabling a unified reconstruction process. Within the LRR block, the INR facilitates the recovery of optimal spatial bases in a low-rank subspace from undersampled measurements. This mechanism effectively exploits redundancy in multi-dimensional k-space data to restore the spatial features of weighted images. These predicted multi-contrast weighted images provide rich structural and quantitative guidance to MLP-driven signal representation in the PMR block, improving the accuracy and robustness of MP-qMRI reconstruction. By integrating dual priors and dual models, LoREIN significantly improves the accuracy and stability of the multi-parametric map reconstruction, especially under high acceleration conditions.

We validated LoREIN on two simulated datasets (with varying contrast information, acceleration factors and temporal dimensions). For comprehensive evaluation, we compared LoREIN against Joint-MAPLE, Zero-DeepSub and SUMMIT on the simulated datasets. Experimental results demonstrate that LoREIN consisently outperforms existing methods. Compared to unrolled-network-based Joint-MAPLE and Zero-DeepSub, LoREIN achieves higher acceleration factors with significantly reduced computational time. Furthermore, LoREIN better exploits redundancy in high-dimensional k-space data than SUMMIT, enabling more stable reconstructions at higher acceleration rates.

The main contributions of our study are:
\begin{enumerate}
\item{LoREIN successfully introduces the first integration of INR with low-rank representation within a dual-model-driven optimization framework, resulting in superior reconstruction for accelerated MP-qMRI.}
\item{LoREIN operates as a novel unsupervised deep learning approach for high-dimensional k-space data reconstruction that delivers superior performance without additional computational cost.}
\item{LoREIN demonstrates exceptional generalization capabilities across diverse acceleration factors, MRI sequences, and image resolutions, ensuring consistent performance regardless of variations in acquisition protocol and temporal dimensionality.}
\end{enumerate}

\section{Related Works}
\label{sec:Related Works}
\subsection{Low-Rank Representation}
In the standard imaging process, the signal $S$ is typically described by the following forward measurement model:
\begin{align}
S =  \mathbf{E}(I) + \varepsilon,\label{bg: eq1}
\end{align}
where $I$ represents the underlying image, $\varepsilon$ is the noise, $\mathbf{E}$ denotes the physical forward operator. Recovering $I$ from the measurement is generally an ill-posed inverse problem and can be formulated as:
\begin{align} 
\arg\min_I \frac{1}{2}\| S - \mathbf{E}(I) \|_2^2 + \lambda \mathcal{R}(I),
\label{bg: eq2}
\end{align}
where the first term imposes data consistency (DC) with measured data, $\mathcal{R}$ is the regularization, characterizing the prior information, and $\lambda$ balances the contributions of these two terms.

However, implementing \eqref{bg: eq2} becomes impractical when the image $I \in \mathbb{R}^{I_1 \times I_2 \times \cdots \times I_q}$ represents a high-dimensional spatiotemporal tensor, due to the computational infeasibility caused by the increasing physical dimensions, also known as the curse of dimensionality \cite{bellman2003dynamic,18_curse_dimension}. To recover the image tensor, a prominent approach is to take advantage of the spatiotemporal correlation to reduce dimensionality. These strong correlations are naturally captured by low-rank representation \cite{zhao2015accelerated,christodoulou2018magnetic,19_LRR_formu}, which has been demonstrated to be a powerful tool for image processing tasks, such as super-resolution and reconstruction \cite{liu2012robust,yaman2019low,xue2021spatial,xue2021spatial}. Theoretically, to capture the spatiotemporal correlation, the image tensor $I$ can be decomposed into the Tucker form as follows:
\begin{align}
I &= \mathcal{G} \times_1 U \times_2 \mathbf{A}^{(2)} \times \cdots \times_{q} \mathbf{A}^{(q)} \notag \\
            &= \left( \mathcal{G} \times_2 \mathbf{A}^{(2)} \times \cdots \times_{q} \mathbf{A}^{(q)} \right) \times_1 U \notag \\
            &= \boldsymbol{\Phi} \times_1 U,\label{bg: eq3}
\end{align}
where the $\times_i$ operator denotes the $i$th mode product, $\mathcal{G}$ is a core tensor, $U$ represents the collection of spatial bases that characterize the underlying spatial patterns, and $\boldsymbol{\Phi}$ is a temporal factor tensor representing all the features associated with temporal bases $\mathbf{A}^{(i)}$. Accordingly, instead of solving \eqref{bg: eq2}, there is a computationally efficient and practical alternative to reconstruct in the following form:
\begin{align} 
\arg\min_{U} \frac{1}{2}\| S - \mathbf{E}(\boldsymbol{\Phi} \times_1 U) \|_2^2 + \lambda \mathcal{R}(U).
\label{bg: eq4}
\end{align} 
By leveraging the low-rank representation, the optimization problem is substantially simplified by solving for a reduced number of spatial bases $U$, rather than the full image tensor $I$. This approach effectively decouples spatiotemporal information, enabling more efficient reconstruction.

\subsection{Implicit Neural Representation}
The concept of INR first emerged in view synthesis with remarkable results \cite{mildenhall2021nerf,rosinol2023nerf}. INR utilizes coordinate grids to approximate continuous signals, offering a flexible approach to model various signals. Although INR is effective for image representation, it suffers from spectral bias, typically resulting in overly smooth reconstructions lacking crucial high-frequency details. Two main approaches have been developed to address this limitation: spatial position encoding techniques (such as Fourier encoding \cite{21_5_fourier_encode} and radial encoding \cite{21_radial_encode}) and the replacement of traditional ReLU activation functions with alternative nonlinear activations \cite{15_encode_INR}.

Recent advances have successfully applied INR to solve inverse problems in medical imaging reconstruction, demonstrating promising results across various modalities including sparse view CT, accelerated MRI, and slice-to-volume reconstruction \cite{chen2025collator,feng2025spatiotemporal,zhang2024subject, chu2025highly,catalan2025unsupervised}. In MRI acceleration specifically, Feng et al. \cite{14_INR_f_mi} proposed IMJENSE, an INR-based approach incorporating sensitivity map estimation for parallel MRI. While effective, this method remains constrained to 2D image reconstruction. Lao et al. \cite{22_lao_singal_model} subsequently developed SUMMIT, extending INR capabilities to 3D MP-qMRI. Although SUMMIT produces promising results for 3D MP-qMRI reconstruction, its performance degrades with highly undersampled data due to insufficient incorporation of prior knowledge about the high-dimensional spatiotemporal k-space data.

\begin{figure*}[!t]
\centerline{\includegraphics[width=0.9\textwidth]{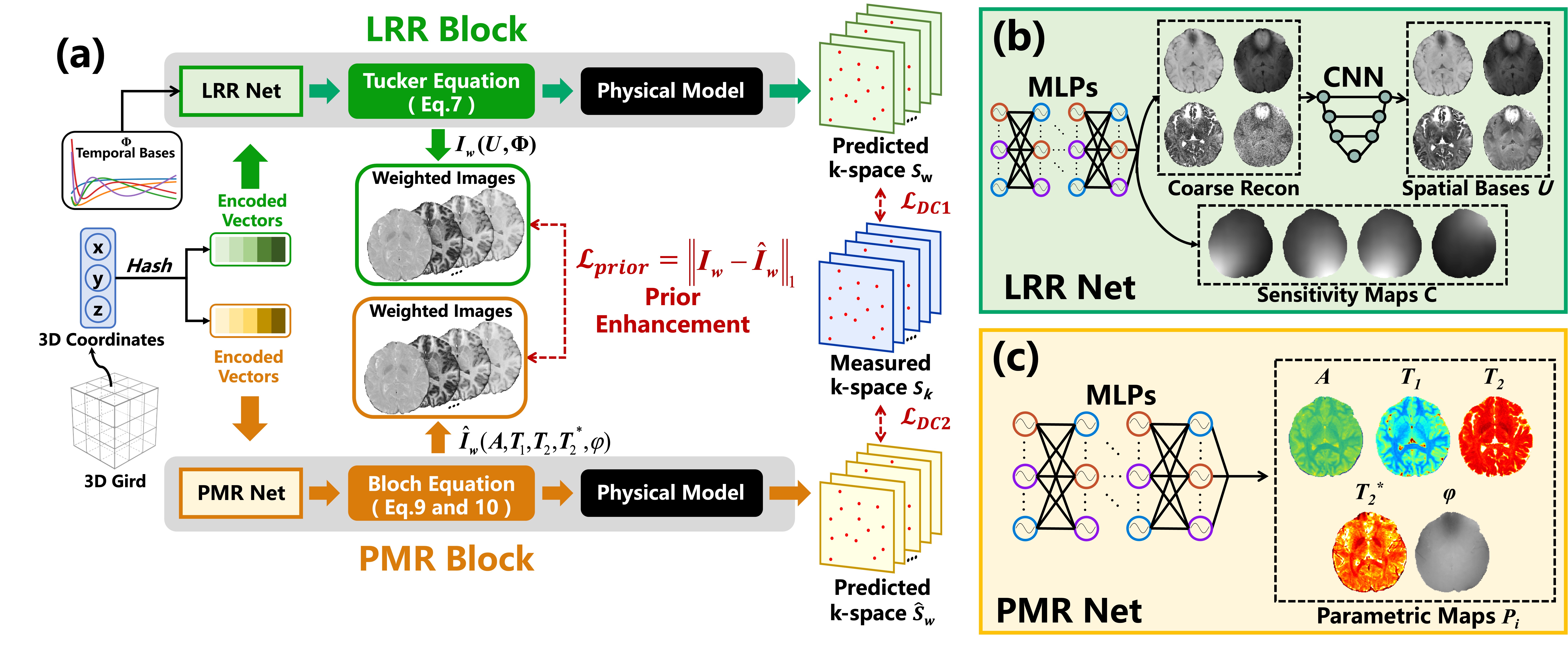}}
\caption{Overview of the proposed method. (a) The 3D coordinates of the voxels are determined by their spatial positions, and the extracted discrete coordinates are processed by hash encoding modules to generate encoded vectors. These encoded vectors are then fed into the LRR net and PMR net, where spatial bases, coil sensitivity maps and parametric maps are produced, respectively. Subsequently, data consistency losses, $\mathcal{L}_{DC1}$ and $\mathcal{L}_{DC2}$, are computed using the forward physical model with the measured k-space data $S_k$. The temporal bases $\boldsymbol\Phi$ are extracted from measured k-space data as well. During reconstruction, weighted images are used to compute the loss $\mathcal{L}_{prior}$ between the LRR block and PMR block, in which the structural prior in the LRR block is leveraged to augment the prediction of parametric maps. (b) In the LRR net, the encoded vectors are passed through MLPs to produce sensitivity maps and coarse spatial bases that are refined into the final spatial bases by a CNN. (c) In the PMR net, the encoded vectors are input into MLPs to generate the desired parametric maps.}
\label{fig_framework}
\end{figure*}

\section{Methods and Experiments}
\label{sec:Methods}
\subsection{Problem Formulation}
\subsubsection{Low-Rank Representation in MP-qMRI Reconstruction}
In MP-qMRI reconstruction, the accurate recovery of weighted images ${I_w}$ from undersampled k-space data is crucial for generating precise quantitative parametric maps of tissue characteristics. Specifically, the undersampled k-space signal $S_k$ is described by the physical forward operator:
\begin{align}
S_k= \mathbf{MFC}I_w,\label{mt: eq00}
\end{align}
where $\mathbf{M}$ denotes the sampling mask, $\mathbf{F}$ represents the Fourier transform, $\mathbf{C}$ are the coil sensitivity maps, and the inverse problem for reconstructing $I_w$ is formulated as:
\begin{align} 
\arg\min_{{I_w}} \frac{1}{2}\| S_k - \mathbf{MFC}I_w\|_2^2 + \lambda \mathcal{R}(I_w).
\label{mt: eq01}
\end{align} 
In MP-qMRI, weighted images are jointly governed by multiple distinct relaxation processes, leading to high spatiotemporal correlations within the images. Therefore, by leveraging low-rank representation theory, weighted images can be effectively decomposed as:
\begin{align} 
I_w=\boldsymbol{\Phi} \times {U},
\label{mt: eq02}
\end{align} 
and consequently, the optimization problem \eqref{mt: eq01} is transformed into the reconstruction of spatial bases ${U}$ as:
\begin{align} 
\arg\min_{U} \frac{1}{2}\| S_k - \mathbf{MFC}(\boldsymbol{\Phi} \times {U})\|_2^2 + \lambda \mathcal{R}(U),
\label{mt: eq03}
\end{align} 
where the temporal bases $\boldsymbol{\Phi}$ are usually extracted from simulated signals or acquired k-space data through singular value decompositions, and the coil sensitivity maps $\mathbf{C}$ are initially estimated from the calibration data using conventional algorithms such as ESPIRiT \cite{uecker2014espirit}. 

Subsequently, based on corresponding signal models, such as the Bloch equation, different quantitative parameter maps are obtained through nonlinear fitting algorithms applied to weighted images $I_w$. Leveraging the low-rank representation for MP-qMRI reconstruction offers a key advantage by significantly reducing the number of unknown variables in the ill-posed inverse problem. This significant reduction in problem complexity enables restoration from highly accelerated high-dimensional k-space data.

\subsubsection{Parameter Quantification in MP-qMRI}
Given the reconstructed weighted images, the estimation of multi-parametric quantitative maps involves performing nonlinear fitting to various signal models, which are intrinsically related to the employed MRI sequences. The signal models primarily include the Bloch equation for different relaxation processes or other sophisticated biophysical models for more diverse tissue characterization. Two signal models are commonly used in practical MP-qMRI sequences. A case in point is the signal model in T2IR-GRE \cite{7_MR_tasking}, where the weighted-image signals are governed by the $T_1$, $T_2$ and $T_2^*$ relaxations of the tissue at each location and described by the Bloch equation:
\begin{align} 
I_w = A \frac{1 - e^{-\frac{T_R}{T_1}}}{1 - e^{-\frac{T_R}{T_1} \cos \alpha}} 
\sin \alpha \cdot e^{-\frac{T_E}{T_2^*}} \cdot e^{j \varphi_e} \cdot \nonumber \\
\left[ 1 + \left( B e^{-\frac{\tau}{T_2}}- 1 \right) 
\left( e^{-\frac{T_R}{T_1} \cos \alpha} \right)^n \right],
\label{mt: eq04}
\end{align}
where $A$ represents the equilibrium amplitude, $B$ represents the effective inversion efficiency independent of $T_2$, $T_R$ is the repetition time, $T_E$ is the echo time, $\tau$ is the T2IR-prep duration, $\alpha$ is the flip angle, $\varphi_e$ denotes the phase at echo $e$, and $n$ is the index of segment within one inversion recovery period. Another typical signal model describes synthetic images that correspond to the joint relaxation processes with parameters $T_1$ and $T_2^*$ \cite{12_joint_maple, 17_VFA_EPTI, zero_deepsub}:
\begin{align} 
I_w = A \frac{1 - e^{-\frac{T_R}{T_1}}}{1 - e^{-\frac{T_R}{T_1} \cos \alpha}} 
\sin \alpha \cdot e^{-\frac{T_E}{T_2^*}} \cdot e^{j \varphi_e}.
\label{mt: eq05}
\end{align}
Although these signal models are relatively accurate in describing the physical processes, the actual fitting procedures face several challenges. The nonlinear fitting processes are computationally expensive, especially for high-resolution 3D qMRI, and highly sensitive to initial parameter settings, which can lead to suboptimal solutions. In addition, these conventional fitting approaches often overlook spatial correlations, leading to nonrobust parameter estimation.

\subsection{MLP and Hash Encoding in INR}
Typically, INR models the target image as a continuous function over a fixed coordinate grid, where MLP serves as a universal function approximator mapping spatial coordinates to intensity values. Meanwhile, hash encoding provides an efficient spatial coordinate embedding strategy by projecting input coordinates into higher-dimensional space, which is subsequently fed into the MLP. Compared to other encoding schemes, hash encoding enables the use of smaller MLPs, leading to significantly faster convergence and improved detail reconstruction \cite{muller2022instant}. The combination of MLP and hash encoding not only reduces GPU memory consumption but also facilitates the representation of high-frequency image components. Furthermore, the inherent continuity property of the INR serves as a powerful regularizer, making the proposed method particularly suitable for MP-qMRI reconstruction from sparsely sampled, high-dimensional k-space data. 

Technically, hash encoding transforms discrete spatial coordinates into $L$ independent and learnable hash tables, each containing $T$ feature vectors of dimensionality $F$. These hash grids represent a set of resolutions in the form of a geometric series, where $N_{min}$ and $b$ are the first term and the ratio of the geometric series. Therefore, these resolutions are $N_{min}, b \times N_{min}, \ldots, b^{L-1} \times N_{min}$. The values of $N_{min}$ and $T$ influence the INR’s ability to capture complex image details, and smaller values often lead to overly smooth outputs. For instance, we introduce a function $f_{\theta}: \mathbb{R}^3 \rightarrow \mathbb{C}$ as a continuous function parameterized by an MLP that incorporates hash encoding. Given a discretized image matrix $P \in \mathbb{C}^{H \times W \times D}$, $(x, y, z) \in \mathbb{R}^3$ represents the related 3D spatial coordinates $( 1 \leq x \leq H, 1 \leq y \leq W, 1 \leq z \leq D )$. Thus, by feeding all the fixed spatial Cartesian coordinates, the image $P$ can be represented as $P_\theta$ through the function $f_\theta$:
\begin{align} 
P_\theta = \begin{bmatrix}
f_\theta(1, 1, 1) & \cdots & f_\theta(1, 1, D) \\
\vdots & \ddots & \vdots \\
f_\theta(H, 1, 1) & \cdots & f_\theta(H, 1, D) \\
\vdots & \ddots & \vdots \\
f_\theta(H, W, 1) & \cdots & f_\theta(H, W, D)
\label{mt:new1}
\end{bmatrix}.
\end{align} 

We stabilize continuous function learning by optimizing hash encoding configurations and MLP architectures based on target image characteristics, thus constraining the solution space and improving reconstruction quality. Specifically, for $A$, $T_1$, $T_2$, and $T_2^*$ maps, the activation function of the last layer in MLPs is an exponential function to ensure positive outputs. For complex-valued images, such as phase maps, coil sensitivity maps, and spatial bases, we employ two separate MLPs to predict the real and imaginary components, respectively. To preserve the full range of complex values, no activation function is applied to the output layer of these MLPs. Moreover, since coil sensitivity maps inherently exhibit smooth characteristics and occupy a low-dimensional space, we control their representation by reducing the resolution and parameter count of the hash encoding. These settings are kept consistent across all datasets.

\subsection{Overall Framework of LoREIN}
Fig. \ref{fig_framework}(a) illustrates the overall framework of LoREIN, which comprises two main components: the Low-Rank Representation (LRR) block and the Parametric Map Reconstruction (PMR) block. Both components utilize hash-encoded 3D spatial coordinates $(x,y,z)$ that are sampled from the image volume grid, and predict relevant outputs through their respective distinct networks (Fig. \ref{fig_framework}(b), Fig. \ref{fig_framework}(c)) and signal models. Within each block, the same forward physical model is used to calculate predicted k-space data at each spatial location. Furthermore, these two blocks are jointly optimized, where the LRR block provides low-rank priors to enhance the reconstruction quality of parametric maps in the PMR block. Ultimately, this integrated framework achieves high-quality reconstruction of MP-qMRI from high-dimensional k-space data.

\begin{figure}[!t]
\centerline{\includegraphics[width= \columnwidth]{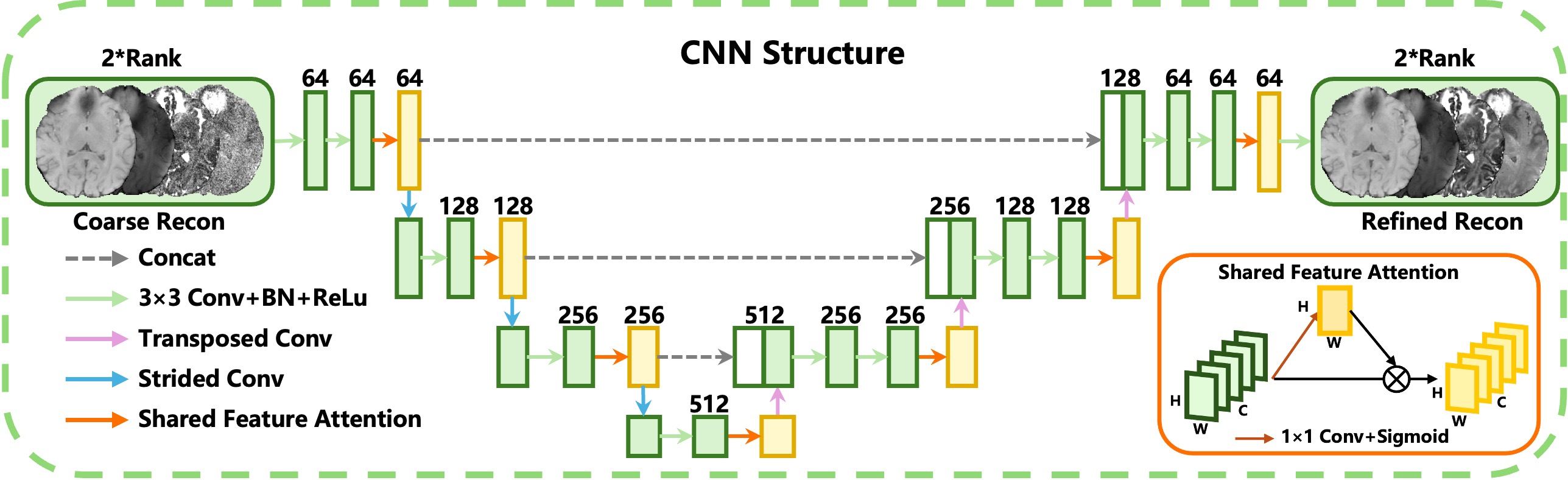}}
\caption{Architecture details of the CNN component in the LRR block. The network processes spatial bases generated by the MLP, with complex-valued inputs separated into real and imaginary channels before concatenation. Rank indicates the number of spatial bases. The network uses strided convolution (stride=2) and transposed convolution for down/upsampling. The inserted shared feature attention module is designed to extract shared structural features across different spatial bases.}
\label{fig_CNN}
\end{figure}

\subsubsection{Low-Rank Representation Block}
INR models the signal as a continuous function over the 3D coordinates $(x,y,z)$, while employing spatial position encoding to mitigate spectral bias. In this work, we extended the INR concept to low-rank representation by designing spatial bases as coordinate-dependent continuous functions. Notably, this integration naturally combines the continuity prior with the low-rank prior. However, components associated with smaller singular values are inherently more challenging to recover in low-rank representation \cite{23_why_cnn}, particularly with highly undersampled data. To alleviate this bias, we introduced a CNN to facilitate the extraction and integration of structural features across spatial bases, thereby improving performance in the restoration of complex spatiotemporal images. Eventually, the proposed approach, which includes a function $f$, parameterized by an MLP incorporating a hash encoding module and a CNN $g$, enables the representation of spatial bases $U$ as:
\begin{align}
f_{\theta} &: \mathbb{R}^{H \times W \times D \times 3} \to \mathbb{C}^{H \times W \times D \times K},  \nonumber\\
g_{\phi} &: \mathbb{C}^{H \times W \times D \times K} \to \mathbb{C}^{H \times W \times D \times K},   \nonumber\\
U &= g_{\phi}(f_{\theta}(x,y,z)), \label{mt: eq3}
\end{align}
where $\theta$ denotes the parameters in the MLP with a hash encoding module, $\phi$ denotes the parameters in the CNN, $H, W, D$ represent the size of the image volume grid and $K$ indicates the number of spatial bases. The detailed structure of the CNN $g$ is shown in Fig. \ref{fig_CNN}. Furthermore, the coil sensitivity maps exhibit smooth characteristics and reside within a low-dimensional space. These attributes render them highly suitable for INR-based function representation, so the framework is capable of representing $\mathbf{C}$ as follows:
\begin{align}
\Tilde{f}_{\Tilde{\theta}} &: \mathbb{R}^{H \times W \times D \times 3} \to \mathbb{C}^{H \times W \times D \times C},  \nonumber\\
\mathbf{C} &= \Tilde{f}_{\Tilde{\theta}}(x,y,z), \label{mt: eq6}
\end{align}
where $\Tilde{\theta}$ indicates the parameters needed to be optimized, and $C$ indicates the number of coils. The final reconstruction problem of spatial bases is written as:
\begin{align}
\arg\min_{\Tilde{\theta}, \phi, \theta} \frac{1}{2}\| S_k - \mathbf{MFC(\Tilde{\theta})}(\boldsymbol{\Phi} \times {U(\phi, \theta)}) \|_2^2. \label{mt: eq7}
\end{align}

\subsubsection{Parametric Map Reconstruction Block}
Leveraging the principle of INR, the multi-parametric quantitative maps are modeled as continuous functions, taking the 3D coordinates as input and producing the corresponding values of different parametric maps at each voxel. We adopted a deep neural network architecture to model these continuous functions implicitly, with each network consisting of a hash encoding module followed by an MLP. Technically, each individual parametric map is represented by a separate function as follows:
\begin{align}
\hat{f_i}_{\hat{\theta_i}} &: \mathbb{R}^{H \times W \times D \times 3} \to \mathbb{R}^{H \times W \times D}, \nonumber \\
P_i &= \hat{f_i}_{\hat{\theta_i}}(x,y,z),\label{mt: eq8}
\end{align}
where $P_i$ indicates different parametric maps, such as $T_1$, $T_2$ and $T_2^*$ maps, $i$ indicates the type of parametric maps. Here, $\hat{\theta_i}$ denotes the parameters of the network for $P_i$ and $\hat{\theta}$ represents the set of all $\hat{\theta_i}$. Depending on the Bloch equation matched to the acquisition sequence, the weighted images can be expressed in terms of the parametric maps. Consequently, the optimization problem \eqref{mt: eq01} can be reformulated as:
\begin{align}
\arg\min_{\hat{\theta}} \frac{1}{2}\| S_k - \mathbf{MFC}\hat{I}_w(\hat{\theta})\|_2^2 +  \sum\limits_{i}{\lambda_i}\mathcal{R}(P_i),
\label{mt: eq9}
\end{align}
where $\hat{I}_w$ is calculated using all parametric maps $P_i$ based on the Bloch equation, and the coil sensitivity maps $\mathbf{C}$ are predicted from the LRR block.

\subsubsection{Prior Enhancement and Loss Functions}
In the LRR block, the low-rank representation is combined with INR, and the parameters of networks are optimized based on the forward physical model. The weighted images calculated from \eqref{mt: eq02} in the LRR block are then used as a structural prior to guide the image signals obtained from parametric maps through the Bloch equation in the PMR block. This cross-block prior enhancement operates through a soft consistency constraint, encouraging the PMR block to generate parametric maps whose forward projections align with the structurally coherent weighted images from the LRR block. Specifically, this mechanism helps preserve anatomical boundaries in the reconstructed parametric maps, while suppressing fluctuations that arise from noisy or incomplete k-space measurements. In this way, the low-rank prior inherent in the LRR is integrated into the PMR, enhancing the expressiveness of the MLPs used to predict parametric maps. Furthermore, to refine the optimization of hash encoding modules and MLPs in the PMR, we applied Weighted Nuclear Norm Minimization (WNNM) \cite{gu2014weighted,22_lao_singal_model} as a regularization term on parametric maps. Eventually, based on the LoREIN framework, the inverse problem for accelerated MP-qMRI reconstruction is formulated as follows:
\begin{align}
\arg\min_{\phi, \theta, \Tilde{\theta}, \hat{\theta}} &\frac{1}{2} \| S_k - \mathbf{MFC(\Tilde{\theta})} (\boldsymbol{\Phi} \times U(\phi, \theta)) \|_2^2 \nonumber \\
& + \frac{1}{2} \| S_k - \mathbf{MFC(\Tilde{\theta})} \hat{I}_w(\hat{\theta}) \|_2^2 \nonumber \\
& + \| \boldsymbol{\Phi} \times U - \hat{I}_w \|_2^2 + \sum\limits_{i}{\lambda_i} \mathcal{R_{WNNM}}(P_i),
\label{mt: end}
\end{align}
and the LoREIN is trained by minimizing the corresponding loss function:
\begin{align}
\mathcal{L}_{tot} = \mathcal{L}_{DC1} + \mathcal{L}_{DC2} + \mathcal{L}_{prior} + \lambda \mathcal{L}_{WNNM},
\end{align}
where $\mathcal{L}_{DC1}$ and $\mathcal{L}_{DC2}$ respectively impose the data consistency in the LRR block and the PMR block, $\mathcal{L}_{WNNM}$ represents the weighted nuclear norm minimization regularization term, and $\mathcal{L}_{prior}$ indicates the prior enhancement from the LRR block to the PMR block.

\subsection{Experimental Data}
In this study, we employed two simulated datasets. Since baseline methods (Joint-MAPLE, Zero-DeepSub, and SUMMIT) were originally developed for different MRI sequences with varying contrast information, sampling strategies, and acceleration factor definitions, we generated the corresponding simulated data and sampling patterns tailored to each method to ensure fair comparison. To facilitate consistent evaluation across all methods, we defined a unified acceleration factor $\text{R}$, formulated as follows:
\begin{align}
\text{R} = \frac{n_{total}} {n_{sampled}},
\label{exp_01}
\end{align}
where $n_{total}$ denotes the total number of k-space data points across all temporal dimensions in the fully sampled case, and $n_{sampled}$ represents the number of k-space data points actually acquired in the undersampled acquisition.
\begin{figure*}[!t]
\centerline{\includegraphics[width=\textwidth]{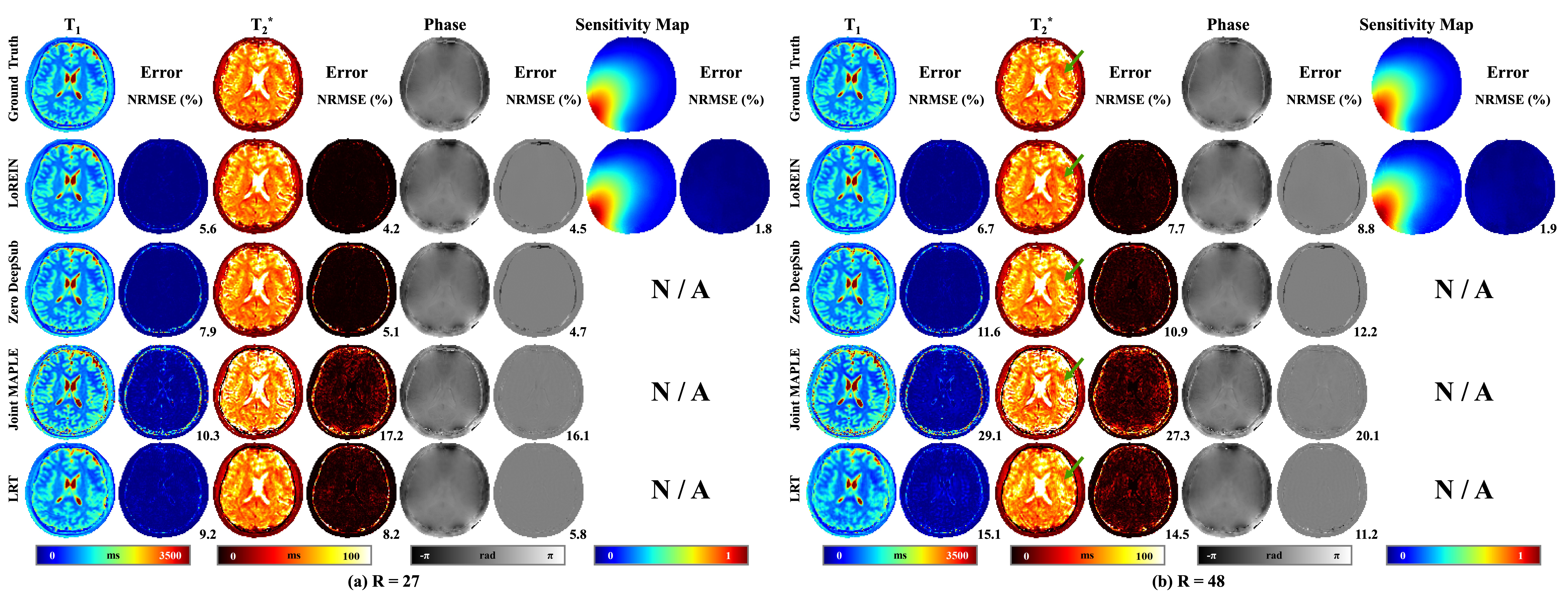}}
\caption{Performance comparison of LoREIN, Zero-DeepSub, Joint-MAPLE, and LRT on simulation dataset 1 at acceleration factors (a) $\text{R}$ = 27 and (b) $\text{R}$ = 48. The first row shows ground truth maps used for simulation. Reconstructed maps from LoREIN, Zero-DeepSub, Joint-MAPLE, and LRT are displayed in the subsequent rows, with corresponding error maps and NRMSE values indicated below. LoREIN yields visually superior images compared to other methods. Sample regions that are more accurately recovered by LoREIN are marked with arrows.}
\label{fig_compare1}
\end{figure*}

\begin{figure}[!t]
\centerline{\includegraphics[width=\columnwidth]{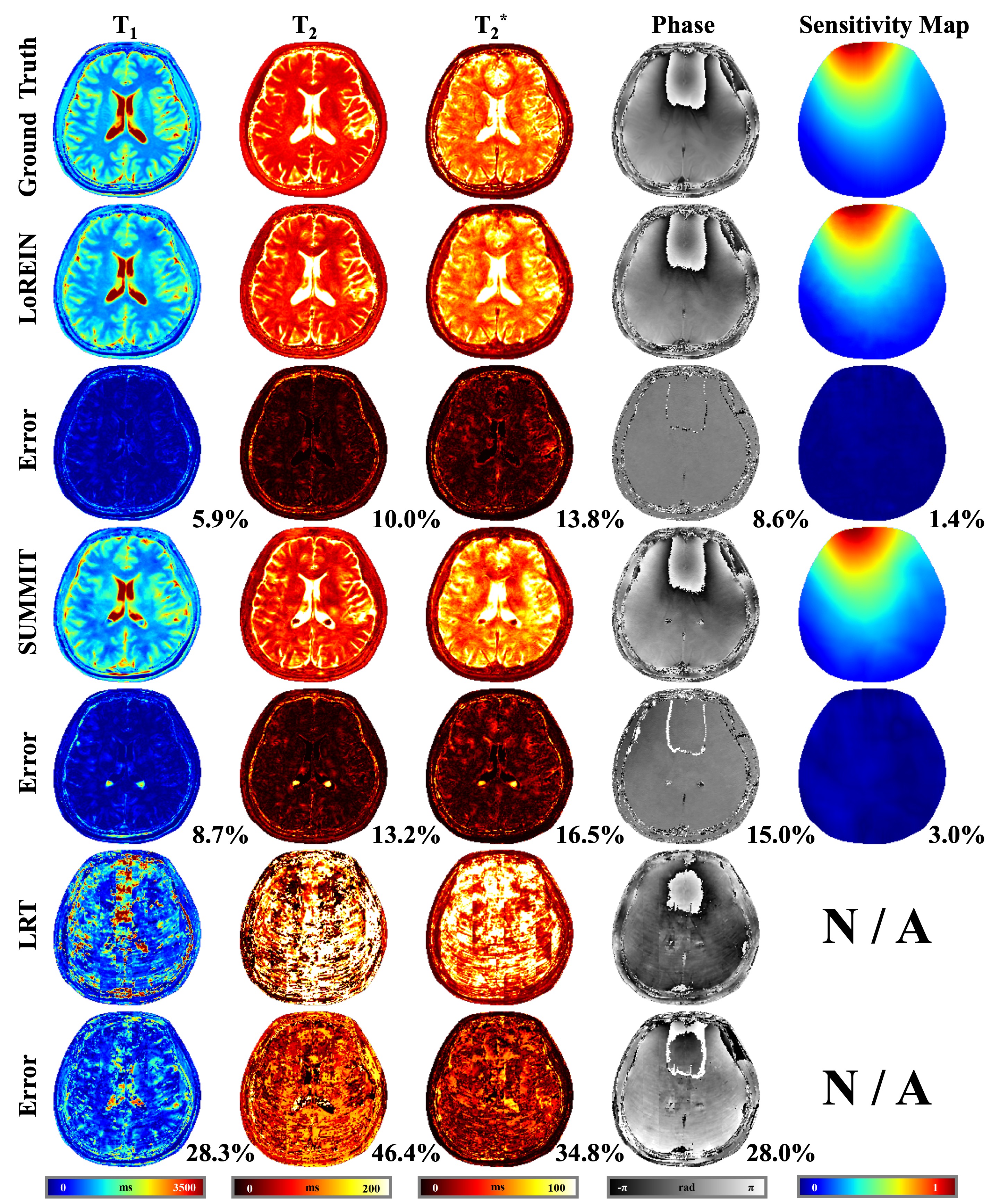}}
\caption{Comparison of LoREIN, SUMMIT, and LRT on simulation dataset 2 at $R$ = 960. The first row shows ground truth maps used for simulation. Reconstructed maps from LoREIN, SUMMIT, and LRT are displayed in the second, fourth, and sixth rows, respectively. Corresponding error maps for each method are shown in the third, fifth, and seventh rows, with NRMSE values indicated below each result.}
\label{fig_compare2}
\end{figure}

\subsubsection{Simulation Data} 
In Simulation Dataset 1, the reference signal model (\ref{mt: eq05}) incorporated the relaxation processes of $T_1$ and $T_2^*$. We retrospectively simulated undersampled k-space data using the following settings: volume size = $112 \times 112 \times 16$, $T_E$ = 2.15 ms to 35.70 ms (3.05 ms interval), $T_R$ = 46 ms, number of flip angles = 5 (5$^\circ$, 10$^\circ$, 20$^\circ$, 30$^\circ$, 40$^\circ$), number of coils = 24, resolution = $1.96 \times 1.96 \times 4$ $\text{mm}^3$ and $\text{R}$ = 12, 20, 27, 48.

For Simulation Dataset 2, we simulated undersampled k-space data based on the T2IR-GRE sequence signal model (\ref{mt: eq04}) using the following settings: volume size = $240 \times 240 \times 14$, $\tau =$ 25, 50, 70, 90 ms, $T_E$ = 4.5, 11.2, 17.9, 24.6 ms, $T_R$ = 30 ms, $\alpha = 10^\circ$, number of segments = 80, number of coils = 14, resolution = $1 \times 1 \times 4$ $\text{mm}^3$ and $\text{R}$ = 160, 240, 320, 480, 640, 960.

\subsection{Implementation Details}
All MLPs in LoREIN employ three hidden layers, each with 64 neurons followed by the ReLU activation function. The number of predicted spatial bases $K$ is set to 15 for all datasets. All hash encoding modules use default hyperparameters \cite{muller2022instant}, except for phase map prediction, where $N_{\min}$ = 1 and $\log_2 T$ = 12. We empirically set weights of the WNNM loss term as $\lambda_1$ = 0.05 for map $A$, $\lambda_2$ = 0.2 for $T_1$ and phase maps, and $\lambda_3$ = 2 for $T_2$ and $T_2^*$ maps. LoREIN is optimized using Adam with an initial learning rate of $1 \times 10^{-3}$. For the simulation data 2, the learning rate decays by half every 20 epochs over 80 total epochs, while for the simulation data 1, it decays by half every 80 epochs over 200 total epochs. To provide more effective guidance, we pre-trained the LRR block for 20 epochs. The implementation used PyTorch 2.0.0 and Python 3.10.11 on a workstation with 40 Intel Xeon Silver 4316 CPUs @ 2.30 GHz, 503 GB RAM, and an NVIDIA GeForce RTX A100 GPU with 80 GB memory.

\subsection{Experiment Design}
To evaluate the proposed method, the simulation datasets were retrospectively undersampled with different acceleration factors. The LRT-based reconstruction method \cite{he2016accelerated}, implemented via the ADMM algorithm with total variation regularization, served as a representative traditional iterative approach in simulation experiments. Specifically, we conducted the following experiments: 

(1) We tested the performance of LoREIN and the baseline methods, Joint-MAPLE and Zero-DeepSub, on simulation data 1 with different acceleration factors. 

(2) We compared LoREIN with SUMMIT on simulation data 2 with large temporal dimensions. Due to strong spatiotemporal correlations in the data, we applied high acceleration factors to highlight the robustness of the method. 

The temporal bases for the simulation data were derived via SVD decomposition of the simulated signals. Unlike LRT, Joint-MAPLE, and Zero-DeepSub, which require pre-estimated coil sensitivity maps via ESPIRiT, LoREIN and SUMMIT simultaneously estimate both coil sensitivity maps and quantitative maps. The normalized root-mean-square error (NRMSE) of the quantitative maps was used as the evaluation metric. In simulation experiments, to mitigate the influence of long relaxation times ($T_1$, $T_2$, $T_2^*$) in the cerebrospinal fluid (CSF) region, we respectively capped these values within 0–3500 ms, 0–200 ms, and 0–100 ms during the NRMSE calculation.

\section{Results}
\label{sec:results}
\subsection{Comparisons on the Simulation Dataset 1}
Fig. \ref{fig_compare1} shows representative brain slice reconstructions from different methods across various acceleration factors. Both LoREIN and Zero-DeepSub demonstrate superior artifact suppression, effectively eliminating the prominent aliasing artifacts observed in LRT and Joint-MAPLE results. This performance advantage is consistently maintained at both $\text{R}$ = 27 (Fig. \ref{fig_compare1}(a)) and $\text{R}$ = 48 (Fig. \ref{fig_compare1}(b)). However, while Zero-DeepSub achieves comparable overall image quality, LoREIN demonstrates superior fine detail preservation, as indicated by the green arrows. Notably, LoREIN consistently achieves the lowest NRMSE across different quantitative maps and effectively estimates sensitivity maps, which cannot be accomplished by Zero-DeepSub, Joint-MAPLE, and LRT methods.

\subsection{Comparisons on the Simulation Dataset 2}
The reconstructed images and the corresponding error maps for LoREIN, SUMMIT, and LRT on simulation dataset 2 are shown in Fig. \ref{fig_compare2}. Compared to LRT, both LoREIN and SUMMIT provide improved image quality and reduced reconstruction errors for $T_1$, $T_2$, $T_2^*$, and phase maps. Additionally, the sensitivity maps estimated by LoREIN and SUMMIT closely match the ground truth, whereas LRT requires ground truth sensitivity maps for reconstruction. However, SUMMIT results still exhibit visible artifacts and inaccurate recovery of image details, which are notably absent in LoREIN reconstructions. The results clearly show that LoREIN demonstrates superior reconstruction performance and achieves the lowest NRMSE errors.

\section{Discussion}
In this study, we proposed LoREIN, a novel zero-shot reconstruction framework for highly accelerated 3D MP-qMRI. By synergistically integrating low-rank and continuity priors while jointly optimizing weighted images and parametric maps, LoREIN achieves superior reconstruction performance. The combination of INR and low-rank representation enables LoREIN to learn optimal spatial bases under highly undersampled conditions, while leveraging structural and quantitative priors from the low-rank model to enhance parametric map accuracy.

We validated the proposed method on retrospective simulation data. The results in Fig. \ref{fig_compare1}, Fig. \ref{fig_compare2} demonstrate the effectiveness of the proposed method for artifact suppression and structural detail preservation. The proposed LoREIN outperforms the baseline methods including the LRT-based method, the state-of-the-art unsupervised method Joint-MAPLE, Zero-DeepSub, and another INR-based method SUMMIT. While Joint-MAPLE directly exploits undersampled data for end-to-end network training, this approach encounters difficulties with high-dimensional k-space data, particularly under extreme acceleration factors, as the ill-posed nature of the inverse problem intensifies, complicating the optimization process. Although Zero-DeepSub exploits low-rank subspaces to reduce the solution space dimensionality of ill-posed problems, it is limited by the low-frequency bias issue of CNN networks, making it challenging to recover fine structural details with increasing sparsity of measurements. Although SUMMIT also leverages the excellent representation capability of INR, it lacks sufficient exploitation of high-dimensional data redundancy, resulting in compromised robustness for high-accelerated k-space data reconstruction. In contrast, LoREIN exploits data redundancy through low-rank representation and utilizes the outstanding signal representation capability of INR, achieving superior reconstruction performance for high-dimensional, highly accelerated MP-qMRI measurements. The results indicate that the proposed reconstruction method holds promise for high-acceleration MP-qMRI reconstruction tasks. Specifically, LoREIN requires approximately 20 minutes for 3D reconstruction on simulation dataset 1, whereas Zero-DeepSub and Joint-MAPLE require at least 1.5-2 hours. Compared to SUMMIT, LoREIN achieves comparable reconstruction time without introducing additional computational overhead. It should be noted that in sequence-based simulation data on T2IR-GRE sequences, we achieved extremely high acceleration factors, primarily due to the extensive temporal dimensions of such sequences (often exceeding thousands of time points). When employing temporally complementary sampling strategies, the number of k-space samples acquired per individual time frame can be substantially reduced. Consequently, when calculating the overall undersampling ratio across all temporal dimensions, the resulting acceleration factors become considerably large.

Several limitations remain in the current method. First, the framework lacks a strategy for handling motion artifacts, which remains an issue in clinical scans, especially for elderly patients and children. Future work could focus on predicting and correcting motion artifacts to enhance clinical robustness. Second, the method has not fully explored correlations between different quantitative maps, such as tissue-specific similarities. Future research will leverage advanced techniques to capture these inter-parametric relationships. Third, the acceleration factors demonstrated in this work represent a careful balance between reconstruction fidelity and acquisition speed; pushing beyond these levels would likely require more comprehensive image priors or architectural innovations to maintain reconstruction quality. Finally, there is potential to improve computational efficiency. Optimizations through model distillation \cite{polino2018model,wu2022tinyvit} or meta-learning approaches \cite{tancik2021learned,NEURIPS2021_61b1fb3f} for network weight initialization could significantly reduce the reconstruction time, enhancing convergence speed and clinical practicality in future iterations.

\section{Conclusion}
In this study, we proposed LoREIN, an innovative  zero-shot reconstruction framework for highly accelerated 3D MP-qMRI. LoREIN learns a continuous implicit functions that directly map spatial coordinates to desired spatial bases and quantitative parameter maps, enabling comprehensive representation of multi-contrast image sequences. The framework uniquely integrates low-rank representation with INR, guided by dual signal models that simultaneously capture spatial coherence and temporal dynamics. Experimental results demonstrate that LoREIN effectively reconstructs fine anatomical details and suppresses noise under high acceleration conditions. Furthermore, LoREIN achieves superior performance across diverse MP-qMRI sequences. We believe that the proposed framework has significant potential to further accelerate high-dimensional (k, t)-space acquisition in the future.

\bibliographystyle{IEEEtran}
\balance
\bibliography{IEEEabrv.bib, Ref.bib}

\end{document}